\documentclass[12pt]{article}
\usepackage[pdftex]{graphicx}

\begin{document}

\def\D3{\overline{\rm D3}}
\def\NS5{\overline{\rm NS5}}

\begin{titlepage}
\setcounter{page}{1} \baselineskip=15.5pt \thispagestyle{empty}
\begin{flushright}
\parbox[t]{2in}{
COLO-HEP-543}
\end{flushright}

\vfil


\begin{center}
{\LARGE Brane/flux annihilation transitions and}\\ {\LARGE nonperturbative moduli stabilization}
\end{center}
\bigskip

\begin{center}
%
{Charles Max Brown and Oliver DeWolfe} \\
\bigskip
\textit{Department of Physics, 390 UCB, 
     University of Colorado,
     Boulder, CO 80309, USA}\\
     \medskip
    \footnotesize{charles.brown@colorado.edu, oliver.dewolfe@colorado.edu}
\end{center}
\bigskip \bigskip \bigskip \bigskip
\begin{center}
{\bf
Abstract} \end{center}

\noindent By extending the calculation of K\"ahler moduli stabilization to account for an embiggened antibrane, we reevaluate brane/flux annihilation in a warped throat with one stabilized K\"ahler modulus.  We find that depending on the relative size of various fluxes three things can occur: the decay process proceeds unhindered, the $\D3$-branes are forbidden to decay classically, or the entire space decompactifies.  Additionally, we show that the K\"ahler modulus receives a contribution from the collective 3-brane tension. This allows for a significant change in compactified volume during the transition and possibly mitigates some fine tuning otherwise required to achieve large volume.

\end{titlepage}
\section{Introduction}
The annihilation of antibranes in the absence of brane partners occurs in a process known as brane/flux annihilation \cite{KPV}.  It is well known that if a collection of D-branes form a stack, the matrix of their positions becomes non-abelian.  In the presence of other fluxes, the famous Myers effect \cite{Myers} occurs, causing the position matrix to become an irreducible representation of SU(2) and effectively generating two new directions for the brane.  Thus a stack of $\D3$-branes ``embiggens"\footnote{This term was applied to the expanded brane in \cite{ABFK1}.  We prefer the descriptive term ``embiggened brane" to the more enigmatic ``giant graviton" which also appears in the literature.} to become a single 5-brane.  The 5-brane can unwrap itself by consuming one unit of $H_3$ flux while traversing a 3-cycle, in the process decaying into a new collection of D3-branes.  This process has been studied in the context of the Klebanov-Strassler solution for a warped deformed conifold throat \cite{KS}, with the expectation that such a throat would be part of a larger compact geometry \cite{Verlinde,GKP}.  Understanding these transitions is essential for building a knowledge of the landscape of string vacua, in particular the history of a universe transitioning within it and the stability of the various solutions; see for example \cite{KKLT}-\cite{Freivogel:2008wm}.

Realistic string vacua must have stabilized moduli, and as is well-known, in type IIB compactifications fluxes generically only stabilize the complex structure moduli and dilaton.  The most well-studied mechanism for K\"ahler moduli stabilization is the introduction of nonperturbative physics hosted on a 4-cycle, either wrapped 7-branes with strong coupling dynamics or instantonic Euclidean 3-branes, as put forward by KKLT \cite{KKLT}.  Such a mechanism can stabilize K\"ahler moduli, but in general introduces a potential for 3-branes as well.  After being described in \cite{Ganor, BHK, GM}, the detailed form of this potential was worked out by Baumann et al.~\cite{Baumann}, and both supersymmetric \cite{DMSU} and nonsupersymmetric \cite{BD} vacua for the combined system of K\"ahler and brane moduli were worked out; for other work using this potential, see \cite{I1}-\cite{CHS}.  Moreover, it was argued in \cite{DMSU, BD} that despite their lack of supersymmetry, $\D3$-branes feel a similar potential in the presence of the nonperturbative physics.

The question thus naturally arises how the nonperturbative moduli stabilization modifies the brane/flux annihilation decay, and therefore how the stability of the landscape of string vacua is affected by nonperturbative moduli stabilization.  It is this question that we study in this paper.  We argue that as the nonperturbative physics is determined to leading order by the backreaction of the brane tension on the warped volume associated to the 4-cycle \cite{Baumann}, the effective tension of the embiggened brane is the relevant quantity.  Calculating the total potential coming from both brane/flux and nonperturbative effects in the case of a single K\"ahler modulus, we show that depending on the parameters of the solution (in particular the overall size of the geometry and the depth of the warped throat, as well as the value of the flux superpotential), a number of different results can occur.

It is possible for the brane/flux potential to dominate, and for the decay process to go through unchanged as described by \cite{KPV}.  Other possibilities exist as well, however.  The nonperturbative potential is negative, and if it dominates, the decay is in general shut down and the world left with negative cosmological constant.  An interplay between the two potentials can exist such that a vacuum that was unstable without the nonperturbative effects can be stabilized near zero vacuum energy.  And finally, vacua with positive cosmological constant may find that they cannot complete the decay, as they experience a classical instability to spontaneous decompactification to ten dimensions during the brane/flux annihilation process.  This can occur even for antibranes that are stable against decay to large volume before the Myers effect takes place.

This paper is organized as follows. In \S2 we establish the setting by reviewing  the physics responsible for brane/flux annihilation in the warped throat.  In \S3 we investigate the non-perturbative physics underlying moduli stabilization and argue for the effect of an embiggened, decaying stack of antibranes on the dynamics.  We also choose a particular embedding for the nonperturbative 7-brane, the ``simplest Kuperstein'' embedding \cite{Kuperstein}, which respects the symmetries of the decay process.  In \S4 we discuss the requirements for volume stabilization, and the beginning- and end-points of the annihilation process.  We then consider the possible potentials along the path of the decay in \S5, and give a summary and example in \S6.  We conclude in \S7.

\bigskip
\section{Review of brane/flux annihilation}

\subsection{Warped throat}

The arena for our investigations is a warped throat of Klebanov-Strassler (KS) type \cite{KS}, attached to a compact bulk geometry; we expect the results to be similar for more intricate warped geometries \cite{ABFK1, ABFK2, EKK}.  In the absence of nonperturbative moduli stabilization and $\D3$-branes, the background is of imaginary self-dual type \cite{GKP}.  The metric takes the form\footnote{We use the conventions of \cite{PS}, which differ from \cite{GKP} by an explicit factor of the string coupling in the definition of the Einstein metric in terms of the string metric, $g^E_{MN} = g_s e^{-\Phi} g^{str}_{MN}$; in these conventions the string and Einstein metrics coincide for the case of constant dilaton.}
\begin{eqnarray}
ds^2=e^{2A(y)}e^{-6u(x)}\tilde{g}_{\mu \nu}dx^\mu dx^\nu+e^{2u(x)}e^{-2A(y)}\tilde{g}_{mn}dy^m dy^n,
\end{eqnarray}
where $e^{A}$ is the warp factor and $e^{2u}$ is the Weyl rescaling required to decouple the overall volume mode from the four-dimensional graviton; we shall always assume $e^{2u}$ as constant over the noncompact space.\footnote{An alternate approach to the realization of the K\"ahler modulus in the ten-dimensional metric has been studied in \cite{GM, Shiu:2008ry, Douglas:2008jx, Frey:2008xw}.  The difference does not affect the effective four-dimensional physics we will study.}  The five-form takes the form \begin{eqnarray}
\label{F5}
\tilde{F}_5\equiv dC_4 - C_2 \wedge H_3 =(1+*){1 \over g_s} [d\alpha(y) \wedge dx^0\wedge dx^1 \wedge dx^2 \wedge dx^3] \,,
\end{eqnarray}
where for imaginary self-dual backgrounds \cite{GM},
\begin{eqnarray}
\alpha(y)= e^{-12u }e^{4A}(y) \,.
\end{eqnarray}
Three-form fluxes $F_3$ and $H_3$ are also generically present in the geometry, obeying the imaginary self-dual condition $*_6 G_3 = i G_3$ with $G_3 \equiv F_3 - \tau H_3$; we will assume for simplicity constant dilaton $e^\Phi = g_s$ and vanishing RR zero-form potential $C_0 = 0$, so $\tau \equiv C_0 + i e^{-\Phi} \to i / g_s$.

We will be focused on the dynamics inside a Klebanov-Strassler throat associated with a particular deformed conifold singularity with 3-cycles $A$, $B$.
The throat is supported by crossed fluxes wrapped on the 3-cycles, 
\begin{eqnarray}
\label{fluxes}
{1\over4 \pi^2}\int_A{F_3}=M\,,\quad {1\over4 \pi^2}\int_B{H_3}=-K\,,
\end{eqnarray}
where we have taken $\alpha' = 1$.
These fluxes stabilize the complex structure modulus $\epsilon$ at an exponentially suppressed value \cite{GKP},
\begin{eqnarray}
\label{epsilon}
\epsilon \approx  e^{-\pi K / g_s M} \,,
\end{eqnarray} 
where $\epsilon$ characterizes the deformation of the conifold through its defining equation \cite{Comments},
\begin{eqnarray}
\label{Conifold}
\sum_{a=1}^4 z_a^2 = \epsilon^2 \,,
\end{eqnarray}
on four complex variables $z_a$.  
In general the deformed conifold's metric can be written in terms of a radial variable, a two-sphere, and a three-sphere; 
the region we are interested in is the tip of the throat, where the two-sphere shrinks to zero size while the three-sphere stays finite.  The unwarped metric at the tip is\footnote{The radial variable here is $\bar{r} \equiv (3/2)^{1/6} \tau$ in terms of the variable $\tau$ of \cite{KS, Herzog}.} \cite{Herzog},
\begin{eqnarray}
\tilde{g}_{mn}dy^m dy^n = \left(2 \over 3\right)^{1/3}\epsilon^{4/3} \left( {1 \over 2} d\bar{r}^2  + d \Omega_3^2  \right)  \,.
\end{eqnarray}
The warp factor, meanwhile, takes the form near the base of the throat \cite{GKP,  Herzog, KKLMMT, GM},
\begin{eqnarray}
e^{-2A} \approx 2^{1/3} g_s M  I_0^{1/2} \epsilon^{-4/3} e^{-2u}  \,,
\end{eqnarray}
up to terms of order ${\cal O}(\bar{r}^2)$, where $I_0 \approx 0.718$ is a pure number.  The total metric at the tip of the throat thus takes the form
\begin{eqnarray}
\label{metric}
ds^2\approx e^{-4u} a_0^2 \,  g_{\mu \nu}dx^\mu dx^\nu+ g_s M b_0^2 \left( {1 \over 2} d\bar{r}^2  + d \psi^2 + \sin^2 \psi d \Omega_2^2 \right) \,,
\end{eqnarray}
where we have for future convenience written the three-sphere metric as a coordinate $\psi$ along with the metric for a two-sphere, and defined the constants
\begin{eqnarray}
a_0^2={\epsilon^{4/3}\over  g_s M (3/2)^{1/3} b_0^2} \,, \quad \quad
b_0^2= \left( 4 \over 3\right)^{1/3} I_0^{1/2} \approx 0.933 \,. 
\end{eqnarray} 
We note that at the tip of the throat, the compact dimensions are independent of both $\epsilon$ and $e^{2u}$; both cancel precisely between the unwarped metric and the warp factor.  Thus the tip of the throat is independent of the Kahler modulus, with its size set purely by $g_s M$.  We shall always take $g_s M \gg 1$ so that the supergravity limit holds at the tip of the throat, as well as $g_s N \equiv g_s M K > (g_s M)^2$ so that $\epsilon \ll 1$.

We assume that this warped throat is part of a larger, compact geometry, though we will make no attempt to characterize the remaining space.  Compactness introduces several issues, the first of which is the need for the $C_4$-tadpole to vanish, which requires
\begin{eqnarray}
\label{tadpole}
{\chi(X)\over24}=Q_3 +{1\over 2\kappa_{10}^2 T_3}\int H_3\wedge F_3 \,,
\end{eqnarray}
where $Q_3$ is the total charge in D3-branes and $\D3$-branes and $\chi(X)$ characterizes the contribution from the 7-branes on the underlying Calabi-Yau 4-fold $X$ in the F-theory picture.
The total space is also characterized by a number of additional complex structure moduli and K\"ahler moduli; the complex structure moduli can be stabilized by three-form fluxes on the associated cycles \cite{OtherFlux, GKP}. 

The K\"ahler moduli, on the other hand, must be stabilized in another fashion; we shall consider the nonperturbative KKLT stabilization \cite{KKLT} associated to 7-branes hosting strong dynamics (or alternately Euclidean D3-branes) wrapping the corresponding 4-cycles.  For simplicity, we will consider just one K\"ahler modulus, associated with the overall volume $e^{4u}$; we shall review the nonperturbative stabilization mechanism shortly.

In principle, the end of the KS throat and the compact geometry beyond provide corrections to the geometry and background fields, even at the bottom of the throat \cite{DKV, AAB, KMS}.  We shall neglect these various corrections, focusing on the influence of only the nonperturbative moduli stabilization effects on the antibrane decay at the bottom of the throat.

\subsection{Life and decay of antibranes}

Moduli-stabilized backgrounds of the type just discussed generally have $AdS$ vacua preserving ${\cal N}=1$ supersymmetry; to obtain $dS$ vacua with no supersymmetry, one mechanism is to place some $\D3$-branes in a the warped throat \cite{KKLT}.
  The action for a D3- or $\D3$-brane in this throat is
\begin{eqnarray}
S_{3\pm} &=& -{\mu_3  \over g_s} \int d^4x \sqrt{-g_4} \pm \mu_3 \int C_4 \,, \\
&=& - {\mu_3 \over g_s} \int  d^4x \sqrt{-\tilde{g}_4} (e^{-12u} e^{4A} \pm  e^{-12u} e^{4A} ) \,, 
\end{eqnarray}
indicating that while a D3-brane feels no force from the background, the $\D3$ experiences a potential, which both gives a positive contribution to the vacuum energy and draws the brane to the bottom of the throat, where by minimizing the warp factor the brane seeks to reduce its energy.

The tip of the throat, however, prevents the antibranes from escaping to zero warp factor, and thus they may settle down into a metastable vacuum.
A further decay mechanism for the antibranes was worked out by Kachru, Pearson and Verlinde (KPV) \cite{KPV} and elaborated on by \cite{DKV}, in which the $p$ $\D3$-branes decay to $M-p$ ordinary D3-branes while changing the flux $K$ by one unit, and this will be the focus of our investigation.

Once our $\D3$-branes reach the tip, the surrounding flux attempts to screen them, lowering the effective charge and raising the effective stress energy, producing a net attractive force and causing the $\D3$-branes to clump \cite{DKV}.  There are also non-perturbative effects, which we will discuss in detail shortly, that draw the clumping $\D3$-branes to the SUSY minimum as if they were D3-branes \cite{BD}.  Once the branes are together, a new set of dynamics becomes available to them\footnote{In contrast to \cite{PRZ}, we assume embiggening only occurs after {\itshape all} $\D3$ branes arrive at the tip of the throat.}: the flux will induce them to puff up or ``embiggen" into a 5-brane wrapping an $S^2$ on the tip via the Myers effect \cite{Myers}.  It is then possible for the antibranes to ``unwrap" themselves across the $S^3$, decaying into ordinary D3-branes in the process.

Here we repeat the results of \cite{KPV} as a review as well as to set our conventions.  We approximate our puffed up $\D3$-branes as one anti-NS5-brane with the appropriate D3-brane charge dissolved in it.  The $\NS5$ wraps a two-sphere on the tip, and we let $\psi$ be the direction normal to the brane on the tip.
The action for our $\NS5$-brane is \cite{PS},
\begin{eqnarray}
S=-{\mu_5\over g_s^2}\int{d^6\xi[-\det(g_{\|})\det(g_{\bot}+2\pi g_s{\cal{F}})]^{1/2}}-\mu_5\int [B_6+ C_4\wedge 2 \pi {\cal{F}}_2],
\end{eqnarray}
where $2 \pi {\cal F}_2 \equiv 2 \pi F_2 - C_2$, and we have separated the metric based on whether it is parallel or perpendicular to the D3-branes which source the $\NS5$-brane. 
The effective charge and tension of $p$ antibranes is encoded in\footnote{Note that the correct D3-brane charge can be realized either with positive $F_2$ on an anti-NS5, or negative $F_2$ on an NS5.  Only the choice we have made leads to the correct potential for antibrane decay, however; the opposite choice can be seen to lead to a potential that gains energy by adding antibrane charge.}
\begin{eqnarray}
\int_{S^2} F_2 = 2 \pi p \,.
\end{eqnarray}
The field strength for $F_3$ can be realized from the potential
\begin{eqnarray}
C_2 = M (\psi - {1 \over 2} \sin 2 \psi) \sin \theta d\theta \wedge d\phi \,,
\end{eqnarray}
where $\theta$ and $\phi$ parameterize the two-sphere, while the six-form potentials may be obtained from \cite{PS}
\begin{eqnarray}
d(B_6 - {i \over g_s} C_6) = {i e^{4A} \over g_s} (*_6 G_3 - i G_3) \wedge dx^0 \wedge dx^1 \wedge dx^2 \wedge dx^3 = 0 \,,
\end{eqnarray}
and thus may be chosen to vanish.\footnote{In \cite{KPV} the WZ contribution to the potential was stated to be obtained from $B_6$, but this is zero; the relevant term actually comes from $C_4 \wedge {\cal F}_2$.}

We can now evaluate several relevant integrals,
\begin{eqnarray}
\int_{S^2}d\theta d\phi \sqrt{\det g_{\bot}} &=& 4 \pi \,b_0^2\, g_s M \sin^2\psi\,, \quad \quad \int_{S^2} 2 \pi {\cal F}_2= 4\pi^2(p  - {M \psi  \over \pi} + {M \over 2 \pi} \sin 2 \psi) \,,\\ 
{1 \over g_s} \int{d^4 x \sqrt{\det g_{\|}}}&=&  \int C_4 \; = \; { a_0^4\over g_s e^{8u}}\int{d^4 x} \sqrt{\tilde{g}_4}\,,
\end{eqnarray}
where recall $\tilde{g}_4$ is the determinant of the noncompact components of the metric stripped of the warp factor and Weyl rescaling; this is the effective 4D metric.
So we can put everything together to obtain the effective four-dimensional action
\begin{eqnarray}
S = - \int d^4x \sqrt{-\tilde{g}_4}  {{\cal C}_0\over e^{8u}}\left[N_3(\psi)\sqrt{1-Z^2e^{4u}\dot{\psi}^2}-  Q_3(\psi)\right] \,,
\end{eqnarray}
where we have defined,\footnote{These functions are related to those in \cite{DKV} by $N_3(\psi) = M V_2(\psi)/\pi$, $Q_3(\psi)= - M U(\psi)/\pi$.}
\begin{eqnarray}
Q_3(\psi)&=& -p +M \left( {\psi \over \pi}  - {1\over2 \pi}\sin 2\psi \right) \,, \quad \quad
N_3 (\psi)= \sqrt{{M^2 b_0^4 \over \pi^2} \sin^4\psi + \left(Q_3(\psi)\right)^2} \,, \\
{\cal C}_0&=& {4 \pi^2 a_0^4 \mu_5 \over g_s}={\mu_3 a_0^4 \over g_s}\,, \quad \quad 
Z^2 ={ g_s M b_0^2\over a_0^2 }\,.
\end{eqnarray}
We have chosen to assume that the only variation is the location of the 5-brane in the $\psi$-direction with time.
For the static case, the effective brane/flux potential is\footnote{The same dependence on the K\"ahler modulus arises in the formalism of \cite{GM, Shiu:2008ry, Douglas:2008jx, Frey:2008xw} in the strongly warped limit.  We thank B.~Underwood for discussions on this point.}
\begin{eqnarray}
\label{Vbf}
V_{bf}(\psi) = {{\cal C}_0\over e^{8u}}(N_3(\psi) - Q_3(\psi))  \,.
\end{eqnarray}
This can readily be interpreted; $N_3$ and $Q_3$ are the effective D3-brane tension and D3-brane charge, in units of a single D3, as the embiggened brane undergoes the transition.    As the brane moves from $\psi=0$ to $\psi=\pi$, the charge $Q_3(\psi)$ varies from $Q_3(\psi = 0) = -p$, corresponding to the initial $p$ $\D3$-branes, to $Q_3(\psi = \pi) = M-p$, the number of ordinary D3's in the final state, thanks to the effect of the coupling to $C_2$.  Meanwhile $N_3(\psi)$ varies between a tension of $p$ branes to a tension of $M-p$, with an additional contribution of the same order of magnitude in the middle stemming from the $\NS5$-brane bubble's finite extent, encoded in the $\sin^4 \psi$ term; this keeps the tension always positive even while the charge is passing through zero.

The $\NS5$-brane is a source for $H_3$, and as it sweeps out the three-sphere, it changes the flux of this field (\ref{fluxes}) from $K \to K-1$.  As a result of this change coupled with the change from $p$ $\D3$-branes to $M-p$ D3-branes, the tadpole constraint (\ref{tadpole}) is preserved, with the only parts changing being the three-brane charge and the fluxes integrated over the cycles associated to our warped throat:
\begin{eqnarray}
- p + KM \quad \to \quad (M-p) + (K-1)M= - p+ KM \,.
\end{eqnarray}
This transition also shrinks the throat slightly as $\epsilon \to e^{\pi/g_s M} \epsilon$, but since $g_s M \gg 1$ the change is small.  Because this decay process involves a removal of both antibranes and flux, it is commonly called {\em brane/flux annihilation}.

\section{Structure of the nonperturbative potential}

Nonperturbative physics on a seven-brane or Euclidean D3-brane wrapping a 4-cycle $\Sigma$ can stabilize the overall volume of a warped compactification; however, this mechanism introduces a potential for D3-branes as well.   In this section, we review these effects and argue for an expression capturing the corresponding potential on an embiggened brane undergoing the brane/flux annihilation decay process.

\subsection{Nonperturbative effects and the embiggened brane}

For an ordinary D3-brane, this physics is determined by a superpotential \cite{Baumann}
\begin{eqnarray}
\label{Wnp}
 W_{np}(Y) = W_0 + A(Y) e^{-a \rho}\,, 
\end{eqnarray} 
with
\begin{eqnarray}
A(Y) \equiv A_0 e^{-\zeta(Y)}  = A_0 f(Y)^{1/n} \,,
\end{eqnarray}
where $W_0$ and $A_0$ are constants that depend on the (already stabilized) values of the complex structure moduli, $Y$ are complex coordinates for the D3-brane, $a$ is the numerical constant $2\pi/n$, $\rho$ is a complex modulus containing the overall volume:
\begin{eqnarray}
\rho \equiv {1 \over 2} (e^{4u} + {1 \over 3} k (Y, \bar{Y}))+ i b \,, 
\end{eqnarray}
$b$ is the axion related to $C_4$ integrated over the corresponding 4-cycle, $k$ is the geometric K\"ahler potential on the space, $n$ is the number of 7-branes/Euclidean D3-branes, and $f$ is their embedding function; that is, the branes sourcing the nonperturbative physics are located at $f=0$.  (Note however that $f$ in the superpotential is a function of the D3-brane location, not the 7-brane location.)  The potential for the coupled system of K\"ahler modulus and 3-brane coordinates is derived from $W$ and the K\"ahler potential $K$:
\begin{eqnarray}
K=-3\log e^{4u}=-3\log(\rho+\overline{\rho}- \gamma k(Y,\overline{Y})/3) \,,
\end{eqnarray}
 in the usual way,
\begin{eqnarray}
\label{Vpot}
V=e^K \left(g^{\alpha\overline{\beta}}D_\alpha WD_{\overline{\beta}}\overline{W}-3 |W|^2 \right), \quad \quad D_{\alpha} W=\partial_{\alpha}W+W\partial_{\alpha}K\,.
\end{eqnarray}
From now on, we will absorb the constant $\gamma \equiv \mu_3 \kappa_{10}^2$ into $k$.

Let us review how the dependence of (\ref{Wnp}) on the brane location was obtained by Baumann et al.~\cite{Baumann}, focusing on the 7-brane case for definiteness.
The nonperturbative physics comes from strong coupling dynamics on a 7-brane, whose strength is controlled by the effective 4D gauge coupling after the wrapped 4-cycle $\Sigma$ is integrated over:
\begin{eqnarray}
{1 \over g^2} = {\mu_3 V_\Sigma^w \over 8 \pi^2} \,.
\end{eqnarray}
Here $V_\Sigma^w$ is the warped volume of $\Sigma$.  The authors of \cite{Baumann} calculated how $V^w_\Sigma$ is modified by the backreaction of the mobile D3-brane tension, and thus how the strength of $W_{np}$ varies with the location of the D3.

The tree-level correction to the warped volume was determined straightforwardly in terms of the perturbation to the warp factor $h \equiv e^{-4A}$ by the brane's tension:
\begin{eqnarray}
\label{warpedV}
\delta V_\Sigma^w = \int_\Sigma d^4 Y \sqrt{g^{ind}} \, \delta h \,.
\end{eqnarray}
This {\em linear} dependence will be of key importance for us.  The effective coupling constant on the 7-brane then enters the superpotential in the usual {\em exponential} form characteristic of instantonic corrections,
\begin{eqnarray}
A(Y) = A_0 f^{1/n}  \propto \exp (- \mu_3 V_\Sigma^w) \,.
\end{eqnarray}
The dependence of the potential on the axion $b \equiv {\rm Im} \ \rho$, in principle determined by the backreaction of $C_4$, was then fixed via holomorphy.

This discussion so far has applied to D3-branes, which in the absence of moduli stabilization feel no force.  In \cite{DMSU}, \cite{BD} it was argued that the nonperturbative potential for an {\em antibrane}, despite the absence of supersymmetry, takes the same form (\ref{Vpot}) with the function $f(Y)$ exchanged for its complex conjugate.  This follows because the {\em tension} of the brane and antibrane are identical, and so the contribution to (\ref{warpedV}), which is determined by a gravitational perturbation, is unchanged at leading order.  

Instead the sign of the $C_4$ fluctuation flips, changing the sign of the contribution that was fixed by holomorphy in the supersymmetric case.  As was noted in \cite{BD}, this affects only a largely insignificant part of the physics, namely 
the fixed value of the axion $b$; other than knowing  that it is fixed at some value, this will play no role for us.

As a result, the nonperturbative effects on a single $\D3$-brane can be determined.  We would like to push this a step further, and determine the effects on a stack of $\D3$-branes as they expand into a 5-brane.  As discussed following equation (\ref{Vbf}), the embiggened brane appears as an effective 3-brane with tension that does not match its charge due to the extra contribution to the tension from the 5-brane's extent over the two-sphere.

However, as we have just reviewed, the tension is far more important for calculating the nonperturbative physics than the charge.  The former determines the coupled equations between the volume modulus and the brane coordinates, while the latter merely shifts the fixed value of the axion $b$, which is not important to our discussion.  It is the tension we need to calculate the perturbation to the warped volume (\ref{warpedV}).

We shall thus focus entirely on the tension, disregarding the details of what happens to $b$.  Since the $\delta V_\Sigma^w$ depends linearly on this tension, we obtain for the warped volume of $\Sigma$ the single-brane result multiplied by $N_3(\psi)$,
\begin{eqnarray}
\delta V_\Sigma^w = N_3(\psi) (\delta V_\Sigma^w)_0 \,.
\end{eqnarray}
This quantity, however, enters the nonperturbative physics in the {\em exponent}, leading to:
\begin{eqnarray}
\label{Afunc}
A(Y) = A_0 f(Y)^{N_3(Y)/n} \,.
\end{eqnarray}
This is the key result, and encodes how the nonperturbative potential is modified by the embiggened brane as its tension changes.

\subsection{7-brane embedding} 

We shall find it useful to pick a particular 7-brane embedding so that we have a concrete value of the function $f$.  Various 7-branes can break the $SO(4)$ isometry of the $S^3$ at the tip in various ways.  Because the embiggened brane expands to fill a set of $S^2$s foliating this $S^3$, it is convenient to pick a 7-brane embedding that preserves the $SO(3) \subset SO(4)$ that acts on this $S^2$.  If one does otherwise, different points on the 5-brane will feel a different potential and it will deform; we choose to avoid this complication. 

A natural choice is the so-called ``simplest Kuperstein" embedding \cite{Kuperstein}:
\begin{eqnarray}
f = z_1 - \mu \,,
\end{eqnarray}
which obviously preserves the $SO(3)$ acting on $z_2$, $z_3$ and $z_4$.  This embedding has vacua for both D3 and $\D3$-branes at $z_1 = \pm \epsilon$ \cite{DMSU}.
To make this embedding compatible with the symmetry of the embiggened brane, we identify $z_1 = - \epsilon \cos \psi$; the brane decay will thus correspond to motion from one vacuum to another, $z = - \epsilon$ to $z = \epsilon$.  

The parameter $\mu$ expresses how far down the throat the 7-brane reaches: $r_{min} \sim |\mu|^{2/3}$.  Strictly speaking $\mu$ is determined by the flux-mediated moduli stabilization that fixes the 7-branes, but as far as we are concerned it is essentially a free parameter.
We can, however, characterize its possible values somewhat: in assuming that this 7-brane has the dominant effect over any other branes or dynamics in the compact geometry, we are assuming it penetrates at least some distance into the throat.
An embedding with $|\mu| > 1$ corresponds to a 7-brane coming only a little ways down the throat, while $|\mu| < 1$ is a ``deep throat" 7-brane.  For the very small value $|\mu|  < \epsilon$, the 7-brane actually passes through the tip \cite{Kuperstein};
we shall in general assume $|\mu| \gg \epsilon$ so the brane doesn't go too deep in the throat, which we expect to be the generic situation.

\subsection{Complex structure and K\"ahler parameters}

There are a number of distinct parameters affecting the physics.  The details of the values of these parameters are in general dependent on properties of the background outside the throat, including the various other three-form fluxes that stabilize the complex structure moduli, 7-brane moduli and dilaton.  In our approximation the parameters are seven in number: the three-form fluxes in the throat $M$ and $K$, the nonperturbative sector parameters $W_0$ and $A_0$, the 7-brane location $\mu$ and the dilaton $g_s$ (all four of which are fixed by unspecified fluxes) and the number of $\D3$-branes $p$.

The two derived parameters that will be most important for us are the complex structure modulus $\epsilon$, which is determined by $K$, $M$ and $g_s$, and the K\"ahler modulus $\rho$, primarily determined by $W_0$ and $A_0$, though as we shall review in the next section,  the other parameters can have an effect as well.  We shall mainly be focused on $\epsilon$ and $\rho$ (as well as the flux superpotential $W_0$) so we review what might constitute ``natural" values for these quantities.

For $\rho$, we have $\rho^{1/4} \sim e^u$ giving the overall length scale of the entire compactification $R$ above the string scale $l_s \equiv \sqrt{\alpha'} = 1$ in our units.  As we shall review shortly, for a standard KKLT compactification we have
\begin{eqnarray}
a \rho = \log \left|A_0 \over W_0\right| \,.
\end{eqnarray}
To trust our geometric description of the model, we require $e^u \gg 1$ ; because of the logarithmic dependence, even relatively modest volumes can lead to $A_0/W_0$ being extremely large.  Thus there is the well-known tension between the desire for large-volume solutions and the fine-tuning of the parameters involved.

The ratio between the scale of physics at the bottom of the warped throat and that at the top is determined by $e^{A_{bott}}/e^{A_{top}} \sim \epsilon^{2/3} e^u/\sqrt{g_s M}$\cite{RS}.
If we take the warped throat to correspond to a scale like the ratio of the electroweak and Planck scales, $M_{EW}/M_{Pl} \sim 10^{-16}$, we obtain
\begin{eqnarray}
\epsilon \sim 10^{-24} (g_s M/ \sqrt{\rho})^{3/4} \sim 10^{-22} - 10^{-26} \,,
\end{eqnarray}
for various reasonable values of $g_s M$ and $\rho$; since according to the definition (\ref{epsilon}) $\log \epsilon \sim -\pi K / g_s M$, it follows that $K$ must then be larger than $g_s M$ by a factor of 50 or so.  As is well-known, a warped throat can generate such an exponential hierarchy with a quite reasonable set of values for the fluxes.  Depending on the physics of the warped throat in question, $\epsilon$ closer to one may also be of interest and does not require significant fine tuning.

\section{Structure of the complete potential}

We now turn to our primary interest: the combination of the last two sections, which is the effect of the nonperturbative moduli stabilization on the antibrane decay process of KPV.  We shall use the expression (\ref{Afunc}) in the potential  (\ref{Vpot}), and compare the resulting potential to that for the brane/flux annihilation (\ref{Vbf}).  We are interested in two degrees of freedom: the collective coordinate $\psi$ standing for the transverse position of the embiggened brane, which will now take the place of the general brane coordinate $Y$, and the volume modulus $\rho$.  
The two pieces of the potential are,
\begin{eqnarray}
V=V_{bf} + V_{np} \,,
\end{eqnarray}
where we can  use the results from \cite{BD} to write the non-perturbative potential at the tip of the KS geometry as
\begin{eqnarray}
\label{Vnp}
V_{np}={a|W_0|^2e^{-8u}\over |\omega|^2}\left({ae^{4u}\over3}+2+\omega+\overline{\omega}+G(\psi)\right)\,.
\end{eqnarray}
Here we have defined 
the ratio between the flux superpotential and nonperturbative superpotential, 
\begin{eqnarray}
\label{omegadef}
\omega\equiv {W_0\over A(\psi)  e^{-a \rho}}\,,
\end{eqnarray}
which plays a key role, as well as the real function of the brane coordinates,
\begin{eqnarray}
\label{Gdef}
G(\psi) \equiv {1\over a}k^{a\bar{b}}\partial_a\zeta\partial_{\bar{b}}\bar{\zeta}\,,
\end{eqnarray}
where we have included the new term from the embiggening (\ref{Afunc}) in 
\begin{eqnarray}
\label{zetadef}
 \zeta \equiv - \log \left(A(\psi) \over A_0\right) =  - {N_3(\psi) \over n} \log f(\psi) \,,
\end{eqnarray}
and the geometric K\"ahler potential can be written as
\begin{eqnarray}
k^{a\bar{b}}\equiv {1 \over Q} ( \epsilon^2 \delta^{a\bar{b}}-z^a \bar{z}^{\bar{b}})\,,  \quad \quad Q \equiv { 2^{1/6}\over   3^{1/3}} \mu_{3}\kappa_4^2 \epsilon^{4/3} \,,
\end{eqnarray}
in terms of the conifold coordinates $z^a$ (\ref{Conifold}), which are in general complex but become real at the tip of the geometry.

Meanwhile we note that $V_{bf}$ (\ref{Vbf}) depends on the volume $\rho$ as,
\begin{eqnarray}
V_{bf} = {{\cal C}_0 \over (\rho + \bar\rho)^2} (N_3(\psi) - Q_3(\psi)) \equiv {D(\psi) \over (\rho + \bar\rho)^2} \,.
\end{eqnarray}
As is well-known, an antibrane alone tends to decompactify the space.  Before turning to the potential as a function of $\psi$, let us examine its behavior as a function of $\rho$.

\subsection{Volume-dependence of $V_{np}$}

As was shown in \cite{BD}, the imaginary part of the $\partial_\rho V$ equations does nothing but require that the ratio of flux to nonperturbative superpotentials $\omega$ is real.  This is realized by adjusting the axion $b \equiv {\rm Im}\ \rho \sim \int_\Sigma C_4$ to cancel out the remaining phase; $b$ appears nowhere else and so is always available to do this.  From now on, we will ignore $b$, take $\omega$ to be real, and for simplicity use $\rho$ simply to refer to what is properly its real part: $\rho = e^{4u}/2$.\footnote{The real part of $\rho$ also includes the geometric K\"ahler potential $k(Y,\bar{Y}$), but this is constant over the tip, and we can absorb this constant into a rescaling of $A_0$. }

For the variation of the potential with respect to the volume, we find
\begin{eqnarray}
\partial_{\rho} V ={-a |W_0|^2\over 24 \rho^3 \omega^2}(3 G (2+2a \rho)+(4+2a\rho)(3+2a\rho+3\omega)) - { D\over 4 \rho^3} \,.
\end{eqnarray}
Equating this to zero, we obtain a quadratic equation for $\omega$.  We will be working consistently within the large-volume limit $\rho \gg 1$, where the solutions are
\begin{eqnarray}
\label{omegasoln}
\omega \approx - {a^2|W_0|^2  \rho \over 2 D}\left( 1 \pm \sqrt{ 1 - {8 \over 3}{ D \over  a |W_0|^2} ( 1  + {3 G \over 2 a \rho}) } \right) \,.
\end{eqnarray}
This solution has a few important properties, the most immediately evident of which is that it has no solution if the parameters of the problem conspire to make the argument of the square root negative.  As we shall see, this provides a key constraint on the stability of the decay process.

The nonperturbative potential (\ref{Vnp}) has two important limits, depending on whether $G$ or $\rho$ is larger.   As we shall describe, when $\rho \gg G$, which we will call the volume-dominated case, the parameter $\omega$ typically has a magnitude of the same order as $\rho$, while for $G \gg \rho$, we call the nonperturbative potential brane-dominated and $|\omega|$ is instead of the same order as $G$.

Let us describe the function $G$ in more detail.  Using (\ref{Gdef}), (\ref{zetadef}) we find
\begin{eqnarray}
G(\psi) = { 1\over a Q} \left(\epsilon^2  \delta^{a \bar{b}} - z^a \bar{z}^{\bar{b}}\right) \partial_a \zeta \partial_{\bar{b}} \bar\zeta \,,
\end{eqnarray}
with 
\begin{eqnarray}
\partial_a \zeta  &=& {z^a \over n z^1} {\partial \over \partial z^1} (N_3 \log f) \,,
\end{eqnarray}
and using $d z^1/ d\psi  = \epsilon \sin \psi$, the reality of the $z^a$ at the tip and $z_a z_a = \epsilon^2 - |z^1|^2 = \epsilon^2 \sin^2 \psi$, we obtain
\begin{eqnarray}
G = {1 \over a n^2 Q} \left(  \log f \, N_3'(\psi) + {\epsilon \sin \psi N_3(\psi) \over f}  \right)^2 \,.
\end{eqnarray}
The $N_3'(\psi)$ term was absent in the case of a fixed number of branes not undergoing an embiggening process.

Let us estimate the size of $G(\psi)$.  At the beginning and end points of the brane decay we have $\sin(\psi = 0,\pi) = N_3'(\psi = 0, \pi)= 0$, and consequently $G=0$ there.  At a generic value of $\psi$, both $N_3$ and $N_3'$ are of order $M$; the latter can cross zero.  Assuming $|\mu| \gg \epsilon$,  these terms are approximately $(M \log \mu) / \epsilon$ and $M/\mu$, and the former term (which is the new term) is generically larger.  (Note that if $N_3'$ crosses zero, it will be smaller in the vicinity of that point.)  If we take the larger term to dominate, we get
\begin{eqnarray}
\label{Gfunc}
G \approx { |\log \mu |^2 \, N_3'(\psi)^2 \over a n^2 Q}  \,.
\end{eqnarray}
Thus estimating the order of magnitude of $G$, we have
\begin{eqnarray}
\label{Gmag}
G &\sim& \epsilon^{-4/3} M^2 \log^2 \mu \,. 
\end{eqnarray}

\subsection{Volume dominance and the endpoints of brane decay}

We consider first the situation where $\rho \gg G$.  This always occurs at least at two very important points: the beginning and end of the brane decay, $\psi = 0, \pi$ where $G=0$ precisely.

The function $\omega$ then has the solution
\begin{eqnarray}
\omega \approx - {a^2|W_0|^2  \rho \over 2 D}\left( 1 - \sqrt{ 1 - {8 \over 3}{ D \over  a |W_0|^2}}\right) \,,
\end{eqnarray}
where for reasons of matching the supersymmetric solution, as we describe momentarily, we have chosen the negative sign on the square root.  It is readily apparent that real solutions do not exist for all values of the parameters; when solutions do exist, it is straightforward to verify that
\begin{eqnarray}
\omega =- {2 \over 3} a \chi \rho \,,
\end{eqnarray}
where $1 \leq \chi \leq 2$ depends on $D/|W_0|^2$.  The nonperturbative potential then has the form
\begin{eqnarray}
V_{np} = - {3 \over 8} {2 \chi - 1 \over \chi^2} {|W_0|^2 \over \rho^3} \,,
\end{eqnarray}
and is always negative.

\subsubsection{Endpoint of brane/flux annihilation}

At the endpoint of the brane/flux annihilation, the system is entirely supersymmetric, consisting of no antibranes, and branes only sitting in the minimum of the nonperturbative potential; moreover, the perturbative potential vanishes, giving $D(\psi = \pi) = 0$.  We then see that, as claimed, the negative sign branch contains the proper smooth limit:
\begin{eqnarray}
\omega_f = - {2 \over 3} a \rho_f\,,
\end{eqnarray}
so $\chi_f = 1$.
Recalling the definition of $\omega$, this reduces to 
\begin{eqnarray}
\left|W_0 \over A(\psi = \pi)\right| e^{a \rho_f} = {2 \over 3} a \rho_f \,,
\end{eqnarray}
the standard KKLT-type relation for the volume modulus.  As usual, a large-volume solution is not generic and can only be obtained by a tuning of the parameters.  Assuming a large volume and taking the logarithm of both sides gives
\begin{eqnarray}
\label{FinVol}
a \rho_f = \log \left|A_0 \over W_0\right| + {M-p \over n} \log |\mu| + \ldots \,,
\end{eqnarray}
where we recalled that $A(\psi) = A_0 f(\psi)^{N_3(\psi)/n}$ and $N_3(\psi = \pi) = M-p$, and took $\mu \gg \epsilon$ so that $f(\psi) \approx \mu$; we have neglected terms of order $\log \rho$.

In a traditional KKLT-type stabilization, it is the smallness of $W_0$ that drives the large size of $\rho$; this can also be achieved with large $A_0$ or a combination of the two.  As we shall see shortly, the magnitude of $W_0$ (but not $A_0$) is constrained by the dynamics.  Note additionally that the D3-brane tension also contributes to the stabilized volume; indeed for natural values of the parameters where $A_0$ and $W_0$ are not far from unity, it can have a significant effect.

\subsubsection{Initial point of brane/flux annihilation}
 
For the point at the beginning of the process where the $\D3$-branes have not yet started to decay, we still have $G=0$ but $D > 0$ in general.  The problem reduces to the standard coupling of the nonperturbative moduli stabilization to an uplifting set of branes explored in KKLT \cite{KKLT}.

Most significantly, we readily see from equation (\ref{omegasoln}) that there is an upper limit on $D$ to permit such a solution:
\begin{eqnarray}
\label{Decompact}
 {3 \over 8} a |W_0|^2 \geq D(\psi =0) \,,
\end{eqnarray}
corresponding to the well-known fact that if the antibrane tension is too strong, it will overwhelm the moduli stabilization and drive spontaneous decompactification \cite{KKLT}.  Thus equation (\ref{Decompact}) corresponds to the {\em decompactification constraint} on the initial vacuum.  Up to numerical factors of order unity, the decompactification constraint for the initial vacuum is
\begin{eqnarray}
|W_0|^2 \geq {p \over M } {\epsilon^{8/3} \over g_s^3 M} \,.
\end{eqnarray}
A large number of antibranes tends to decompactify the space, but this is strongly mitigated by a small complex structure parameter $\epsilon$ reducing their effective tension.  

We will assume this bound is satisfied so that such a vacuum exists.   We then find 
\begin{eqnarray}
\omega_i =- {2 \over 3} a \chi_i  \rho_i \,,
\end{eqnarray}
where as before $\chi_i$ depends on $D_i/a |W_0|^2$ and for permitted values, varies merely between $1 \leq \chi \leq 2$.  We thus have for the initial volume,
\begin{eqnarray}
\label{BegVol}
a \rho_i = \log \left|A_0 \over W_0\right| + {p \over n} \log |\mu| + \ldots \,,
\end{eqnarray}
where we dropped $\log \chi$ as being insignificant.

Notice that if we assume that the large volume is primarily due to $W_0$ being very small as $\rho \sim \log |1/W_0|$, we can easily run afoul of the decompactification constraint.  The ratio $A_0/W_0$, however, is unfixed and so in principle can be tuned to produce any volume despite the constraint; a large $A_0$ is in general required.

Also note that the {\em difference} between final and initial volumes is
\begin{eqnarray}
\Delta \rho \approx {M - 2p \over 2\pi} \log |\mu| \,.
\end{eqnarray}
This is a key result: in order for the volume not to change substantially during the brane/flux annihilation process, we must have
\begin{eqnarray}
\label{Rholimit}
\rho \gg  \Big| {(M-2p)\over2\pi} \log |\mu| \Big|\,.
\end{eqnarray}
Since $M$ may be of order hundreds or more, as we require $g_s M \gg 1$ with $g_s$ small, this volume change becomes quite relevant for values of $\rho$ that are not too large.  This is our second key result: a brane/flux decay with only modestly large volume will in general change its volume during the course of the decay process.

Before moving on to a study of the middle of the decay, let us consider the initial cosmological constant in the system.  This comes from an interplay between $V_{bf}$, which is positive, and $V_{np}$, which is negative:
\begin{eqnarray}
\Lambda = V_{bf,i} + V_{np,i} = |V_{np,i}| (\lambda - 1) \,,
\end{eqnarray}
where we have defined
\begin{eqnarray}
\label{CosmologicalConst}
\lambda \equiv |V_{bf, i}|/|V_{np, i}| \sim {D_i \rho_i \over |W_0|^2} \sim {p \over M} {\epsilon^{8/3} \over g_s^3 M} {\rho_i \over |W_0|^2} \,.
\end{eqnarray}
Evidently, $\lambda > 1$ leads to an initial metastable state with a positive cosmological constant, while $\lambda < 1$ implies a negative cosmological constant.  A realistic present-day vacuum energy,
in which case the brane annihilation process would correspond to the decay of our world, would require $\lambda$ extremely close to one,
\begin{eqnarray}
|W_0|^2 \sim D_i \rho_i \,,
\end{eqnarray}
which, it is nice to verify, satisfies the decompactification constraint (\ref{Decompact}) with a factor of $\rho$ to spare.  This constraint prevents existence of metastable vacua with larger positive cosmological constant, characterized by $\lambda \sim \rho$ or greater.

\subsubsection{General volume dominance}

During the brane decay $0 <\psi <\pi$, it is possible for $G$ to grow and overwhelm the volume, changing the form of the potential.  Given the estimate for $G$ (\ref{Gmag}), this will not happen and the volume will continue to dominate the potential for
\begin{eqnarray}
\rho \gg \epsilon^{-4/3} M^2 \log^2 \mu \,.
\end{eqnarray}
For the modulus $\epsilon$ chosen to solve the hierarchy problem, we expect the right-hand-side to be in excess of $10^{32}$.  Thus volume-domination of the nonperturbative potential persists only if the volume is very large indeed; we need length scales $10^8$ times the string scale for universal volume-domination, bringing us into the realm of large extra dimension models.  (The fine-tuning of the parameters $W_0$ and $A_0$ required to obtain such a large volume is truly extraordinary, as well.)  We shall therefore largely focus on the brane-dominated nonperturbative potential.

\subsection{Brane dominance}

Away from the beginning and the end of the decay, we will have $G \gg \rho$ if 
\begin{eqnarray}
\epsilon^{-4/3} M^2 \log^2 \mu \gg \rho  \,,
\end{eqnarray}
and as just argued we expect this situation to be the easier to realize.  The solution (\ref{omegasoln}) for $\omega$ becomes
\begin{eqnarray}
\label{Gomegasoln}
\omega \approx - {a^2|W_0|^2  \rho \over 4 D}\left( 1 - \sqrt{ 1 - {4  D G \over  a^2 |W_0|^2 \rho} } \right) \,,
\end{eqnarray}
which takes the value
\begin{eqnarray}
\omega = - \chi' G \,,
\end{eqnarray}
with $1 \leq \chi' \leq 2$ when a solution is possible.  The nonperturbative potential in this limit takes the form
\begin{eqnarray}
\label{Vnpbrane}
V_{np} \approx  - { a (2 \chi'(\psi) -1) |W_0|^2 \over 4 \chi'(\psi)^2 \rho^2 G(\psi)} \,.
\end{eqnarray}
In the next section, we shall describe the consequences of the nonperturbative potential on the decay of the embiggened antibranes.

\subsection{Validity of the nonperturbative expansion}

The nonperturbative term going like $e^{-a \rho}$ in (\ref{Wnp}) is in principle only the leading term in an expansion of exponentials in $\rho$; it is only valid to neglect other terms when the leading term is small.  At large volume $e^{-a \rho}$ is always small, but one may wonder whether $A(\psi) e^{-a \rho}$ might still be large, thus invalidating the expansion.

Using the definition (\ref{omegadef}) of $\omega$, the size of the nonperturbative term is
\begin{eqnarray}
\label{Validity}
A(\psi) e^{-a \rho} = { W_0 \over \omega} \,,
\end{eqnarray}
and this furnishes an upper bound on $|W_0|$ to be less than $\omega$, which we have now shown goes like the larger of $\rho$ and $G$.  Since $|W_0|$ is generally taken small to support large volume solutions, this is not in general a difficult constraint to satisfy.

\section{Moduli-stabilized brane/flux transitions}

In what follows, we will study the $\psi$-dependence of the total potential with (\ref{omegasoln}) enforcing that it sits at an extremum in the $\rho$-direction (if such an extremum exists).  Thus we are assuming that the $\partial_\rho V = 0$ equation, which held at $\psi = 0$ to define the initial vacuum, remains satisfied and adjusts the system accordingly as the the brane rolls down a slope in $\psi$.  We expect that an analysis involving both $\rho$ and $\psi$ as dynamical fields, beyond the scope of this paper, would reflect the same phenomena we uncover here.

As we have already seen, several interesting things can occur during the decay of the embiggened brane under the influence of the nonperturbative potential.  We have reviewed how there is a limit on the initial (positive) cosmological constant to have a metastable vacuum at all; in the next subsection, we shall describe how even if this constraint is satisfied, certain positive cosmological constant models violate the related bound during the decay, and thus decompactify.  We shall then discuss how for cosmological constants near zero it is possible for the nonperturbative potential to stabilize an otherwise perturbatively unstable solution, potentially saving a supersymmetry-breaking universe like our own from quick decay.  Total dominance of the nonperturbative potential, however, occurs only for negative initial cosmological constants.  Finally, we note how even if the transition completes, the overall volume can be changed.  In the next section, we present an example illustrating these possibilities.

\subsection{Spontaneous decompactification}

For the cases where $G \gg \rho$ during the decay (or even $G \sim \rho$), we can run into trouble satisfying the $\partial_\rho V = 0$ equation (\ref{omegasoln}).  We assumed that $|W_0|$ took a value such that a solution existed before the decay, requiring (\ref{Decompact}); even so, we can run into difficulties in the middle of the decay process.

The decompactification constraint for $G > \rho$ is modified: solutions to (\ref{Gomegasoln}) require 
\begin{eqnarray}
\label{DecompactificationLimit}
a^2 |W_0|^2 \rho > 4 D(\psi) G(\psi)  \,,
\end{eqnarray}
and using the estimate
\begin{eqnarray}
D \approx {\epsilon^{8/3} \over g_s^3 M} \,,
\end{eqnarray}
we can parameterize the constraint as approximately,
\begin{eqnarray}
\label{Decompact2}
|W_0|^2  >  {\epsilon^{4/3} M \log^2 \mu \over g_s^3 \rho} \,.
\end{eqnarray}
It is entirely possible for $|W_0|$ to be such that the initial metastable vacuum for the antibranes exists, but the constraint is then violated during the course of the decay (for an example, see figure \ref{Figomega} in the next section).  We can parameterize this in terms of the initial cosmological constant parameter $\lambda$; we find
\begin{eqnarray}
\lambda < {p \over M} {\rho^2 \over G} \,,
\end{eqnarray}
while a metastable vacuum at the outset only required $\lambda < \rho$.

The inability to satisfy the equation (\ref{omegasoln}) at some point during the decay means the potential in the $\rho$-direction no longer has a local minimum; the potential has ``opened up" to a slope leading down towards large volume.  Although in principle one would need to solve the full dynamical coupled system of $\psi$ and $\rho$ including kinetic terms to obtain the full motion, we generically expect that once the system finds itself without a minimum in the $\rho$-direction, it will roll to large volume.

Thus it is is possible for the brane/flux annihilation process to spontaneously decompactify the otherwise metastable system.  The decay of the embiggened brane presumably completes, but the system is no longer of phenomenological interest; it has decayed to ten-dimensional flat space.  Metastable vacua with large initial cosmological constant are the most in danger of this.

Can this occur for an initial state with vacuum energy near zero?  This coincides with $\lambda =1$, implying for stability
\begin{eqnarray}
\epsilon^{-4/3}  {M \over p} M^2 \log^2 \mu< \rho^2  \,,
\end{eqnarray}
where we used (\ref{Gmag}) to estimate $G$.

This requires a substantial volume. For example, for the case of the throat solving the hierarchy problem, we need something of the order $\rho > 10^{16}$, or length scale $10^4$ times the string scale.  While not as large a volume as was necessary for the nonperturbative potential to become volume-dominated, this is still a substantial volume representing a tremendous tuning of $A_0$ relative to $W_0$.  

Thus we see that a set-up with a realistic cosmological constant may spontaneously decompactify before the brane decay can complete for a modest compact volume.  In general, a deep throat encourages spontaneous decompactification, while a large volume discourages it; the former is much more generic in the space of parameters.

\subsection{Nonperturbative dominance}

Assuming the system satisfies the decompactification constraint throughout  its evolution, the next question to ask is whether the nonperturbative potential is smaller than, similar to or much larger than the perturbative potential.  In the more generic case of $ G \gg \rho$, we find
\begin{eqnarray}
{|V_{np}| \over| V_{bf}|} &\sim& {|W_0|^2  \over G D} \,, \\
&\sim& {1 \over \rho} {|W_0|^2 \over |W_0^{min}|^2} \,,
\end{eqnarray}
where $ |W_0^{min}|^2$ is defined by the minimum allowed value for the decompactification constraint (\ref{Decompact2}).

We readily see that if $W_0$ is near the smallest value allowed to prevent spontaneous decompactification, the nonperturbative potential is suppressed relative to the brane/flux potential by a factor of the large volume.  In this region, then, the brane decay potential is unmodified from the original study of \cite{KPV}, and the decay goes through as described there.  Thus, it is entirely possible for brane/flux transitions to proceed unmodified by the nonperturbative physics.

In terms of the initial cosmological constant parameter $\lambda$, we have
\begin{eqnarray}
{|V_{np}| \over| V_{bf}|} \sim {p\over M} {\rho \over G} {1 \over \lambda} \,,
\end{eqnarray}
demonstrating that the $\lambda \gg 1$  systems with large cosmological constant, the ones in danger of decompactifying, are always dominated by the brane/flux potential during their evolution.  Thus positive cosmological constant metastable states, if they do not decompactify, tend to decay in the original way described by KPV.

The vacua where $V_{np}$ clearly dominates have $\lambda \ll 1$, that is, they have negative cosmological constant.  These are less interesting phenomenologically, since our universe has positive vacuum energy and any hypothetical states in the early universe that decayed to our vacuum would have had larger energy still.
$V_{np}$ for $G \gg \rho$ has its $\psi$-dependence primarily from $-1/\rho^2 G(\psi)$, except at the endpoints, where it becomes $-1/\rho^3$.  So, $V_{np}$ is most negative near $\psi = 0$ and $\psi = \pi$, and has a less negative ``bump" in the middle, of size $\rho/G$ closer to zero; thus $V_{np}$ tends to rise in the middle of the annihilation process, shutting down the decay and stabilizing the system at small $\psi$ if it is dominant (see for example figure \ref{Fig2} in section~\ref{SummarySec}).

It is, however, possible for models near $\lambda =1$, a realistic cosmological constant, to have the two potentials similar sizes, $|V_{np}| \sim |V_{bf}|$.   Even if the brane/flux potential would lead to unobstructed decay, the ``bump" from the nonperturbative potential can in this case provide a stabilizing barrier.  Thus it is possible for a universe with vacuum energy similar to ours which was perturbatively unstable from $V_{bf}$ alone, to be stabilized perturbatively by the nonperturbative potential, as in figures \ref{Fig06} and \ref{Fig05}. 

Inflation in the context of the brane/flux annihilation decay was studied in \cite{DKV} (see also \cite{PRZ}), and when the brane/flux potential dominates the story from there will be unchanged.  Inflation obviously will not occur for nonperturbative-dominant cases, which have negative vacuum energy; it is possible that the regime where both potentials contribute could lead to an enhanced flat direction if they are tuned just right, but the caveats of the difficulty of such a tuning outlined in \cite{DKV} still remain.

\subsection{Finite volume change}

While the brane decay is proceeding, the value in the $\rho$-direction that minimizes the potential may vary.  The analogous equation to (\ref{FinVol}) and (\ref{BegVol}) for any point along $\psi$ is
\begin{eqnarray}
a \rho(\psi) =  \log \left|A_0 \over W_0\right| + {N_3(\psi) \over n} \log |\mu| + \ldots  \,,
\end{eqnarray}
where $\ldots$ contains $\log \rho$ or $\log G$, depending on which is larger.  The corresponding volume change associated to the second term is of the order,
\begin{eqnarray}
\Delta \rho \sim M \log \mu\,.
\end{eqnarray}
It is useful to note that $G$ is of the order (\ref{Gmag}),
\begin{eqnarray}
G \sim \epsilon^{-4/3} (\Delta \rho)^2 \,,
\end{eqnarray}
which implies that in the volume-dominated case of $\rho \gg G$, 
\begin{eqnarray}
{\Delta \rho \over \rho} \ll {\epsilon^{2/3} \over \rho^{1/2}} \,,
\end{eqnarray}
and no significant volume change occurs.
 However, in the more generic case of brane-dominance, it is possible for such a change to occur.  If $\log |A_0/W_0|$ is only of order hundreds or thousands, it will generically be the case that the volume change will be significant, even if $\log |\mu|$ is of order one.
 
Note that the {\em sign} of the volume change is determined by $\log |\mu|$: a 7-brane only going a little way into the throat will have $\log |\mu| > 0$  and hence $\Delta \rho > 0$, while a 7-brane penetrating far into the throat will produce $\Delta \rho < 0$.  Since we expect the former to be more common, volume increase is more generic, but either is possible as in figure~\ref{Figrho}.

We also note the possibility that the volume may be dominated by the flux at all points: $N_3(\psi) \log |\mu|  \gg \log |A_0/W_0|$ for all $\psi$.  The volume would then simply be
\begin{eqnarray}
a \rho(\psi) \approx {N_3(\psi) \over n} \log |\mu|  \,,
\end{eqnarray}
and would grow by a factor of $M/p$ during the course of the decay.  ($\log |\mu| < 0$ in this case would lead to negative volume, outside the bounds of our large-volume approximation.)
Although this is a different scenario than the one usually envisioned in a KKLT-type model, it may actually be somewhat generic, as large $M \log \mu$ is substantially more natural than large $\log |A_0/W_0|$ due to the logarithm.   However, we note that in this case $G \sim \epsilon^{-4/3} \rho^2$, implying $G \gg \rho^2$; one can then show that satisfying the decompactifcation constraint  (\ref{DecompactificationLimit}) implies that $\lambda < 1$ in  (\ref{CosmologicalConst}), leading to an unphysical  negative cosmological constant at the beginning and end of the transition.

\section{Summary and Example}
\label{SummarySec}

  In summary, we have found the following broad classes of dynamics for the moduli-stabilized embiggened brane system:

\begin{itemize}

\item{Metastable vacua with large positive cosmological constant can spontaneously decompactify during the decay, returning to ten-dimensional flat space.  If they do not decompactify, the brane/flux annihilation proceeds as originally described.}

\item{Vacua with vacuum energy near zero can be unstable to decompactification, but can also be stabilized against perturbative decay by the nonperturbative physics.}

\item{Vacua with negative cosmological constant are dominated by the nonperturbative potential, and the decay is shut down.}

\item{If the initial volume is only modestly large, it can grow or shrink during the course of a successful brane decay, with the sign of the change determined by how deep the 7-brane hosting nonperturbative physics penetrates into the throat.}

\end{itemize}

\begin{figure}
\begin{center}
\includegraphics[width=0.6\textwidth]
{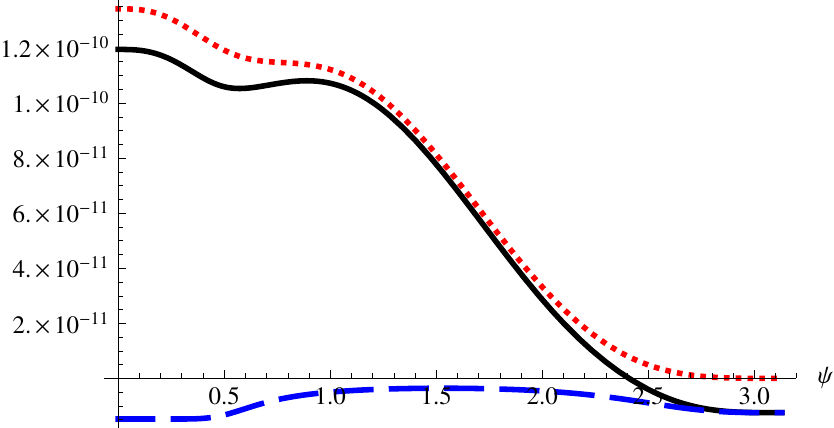}
\caption{The full potential (black, solid) and its components $V_{np}$ (blue, dashed) and $V_{bf}$ (red, dotted) for $|W_0| = 0.2$, dominated by the brane/flux potential.
\label{Fig02}}
\end{center}
\end{figure}

For an example, consider the following choices of parameters:
\begin{eqnarray}
\epsilon = 0.05 \,, \quad \log \left| A_0 \over W_0 \right| = 1000 \,, \quad g_s = {1 \over 10} \,, \quad M = 100 \,, \quad \mu = 2 \,, \quad {p \over M} = 0.08\,,
\end{eqnarray}
where we will allow $W_0$ to vary in the vicinity of unity; the nonperturbative term (\ref{Validity}) is always small.  Note that this is not a particularly realistic model for solving the hierarchy problem, as the throat is extremely shallow; we chose it for computational convenience rather than realism.

\begin{figure}
\begin{center}
\includegraphics[width=0.6\textwidth]
{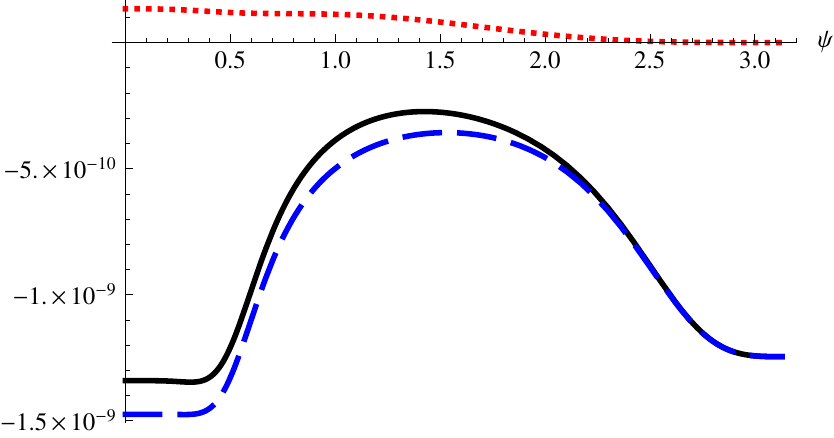}
\caption{The full potential (black, solid) and its components $V_{np}$ (blue, dashed) and $V_{bf}$ (red, dotted)  for $|W_0| = 2$, dominated by the nonperturbative potential.
\label{Fig2}}
\end{center}
\end{figure}

\begin{figure}
\begin{center}
\includegraphics[width=0.6\textwidth]
{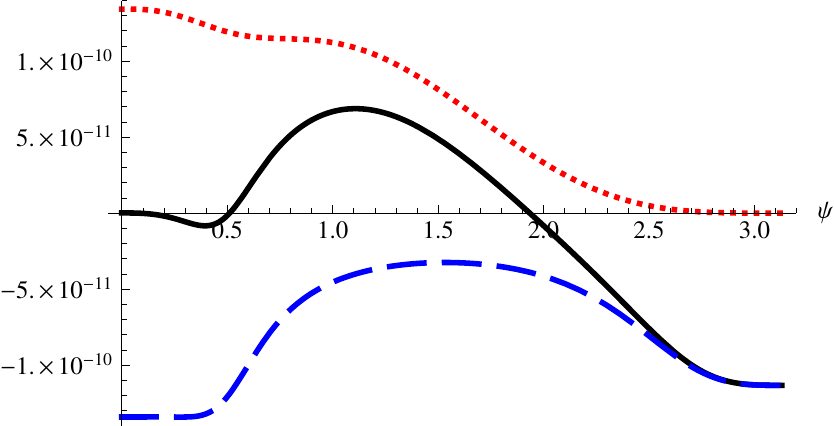}
\caption{The full potential (black, solid) and its components $V_{np}$ (blue, dashed) and $V_{bf}$ (red, dotted)  for $|W_0| = 0.603$, with initial vacuum energy close to zero.
\label{Fig06}}
\end{center}
\end{figure}

\begin{figure}
\begin{center}
\includegraphics[width=0.6\textwidth]
{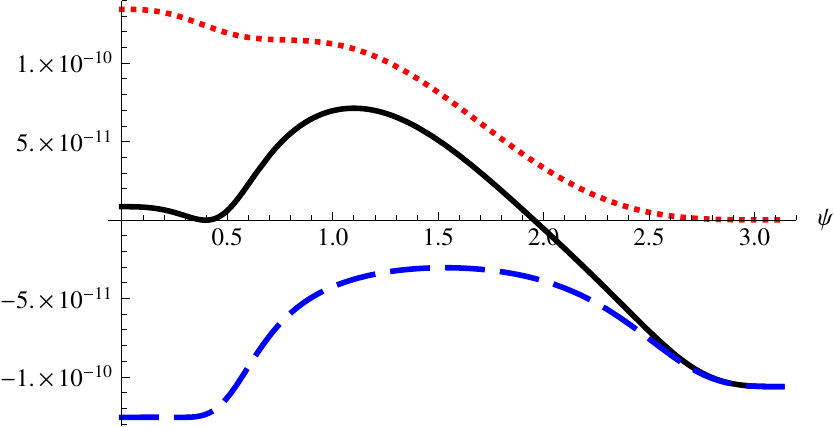}
\caption{The full potential (black, solid) and its components $V_{np}$ (blue, dashed) and $V_{bf}$ (red, dotted) for $|W_0| = 0.584$, with metastable vacuum energy close to zero.
\label{Fig05}}
\end{center}
\end{figure}

For $|W_0| < 0.2$, the brane/flux potential is dominant; this is plotted in figure~\ref{Fig02}.
Meanwhile for $|W_0| > 2$, the total potential is dominated by the nonperturbative contribution, as plotted in figure~\ref{Fig2}.  In the former case the initial vacuum energy is positive, and the decay process proceeds unobstructed, while in the latter case the vacuum energy is negative and the decay is obstructed perturbatively.  We choose to plot values where the smaller potential still makes some contribution for visual interest; as $W_0$ moves further away from the cross-over region, one potential or the other dominates completely.

For intermediate values both potentials contribute; in particular, in the vicinity of $|W_0| = 0.60325$ the initial cosmological constant is tiny as resulting from a near-perfect cancelation of the two contributions to the potential; the system embiggens only slightly to a metastable vacuum with a negative cosmological constant, figure~\ref{Fig06}.  Perhaps of more interest is $|W_0| = 0.5837$ (figure~\ref{Fig05}), where the brane grows slightly and drops into a metastable vacuum with cosmological constant near zero.

\begin{figure}
\begin{center}
\includegraphics[width=0.4\textwidth]
{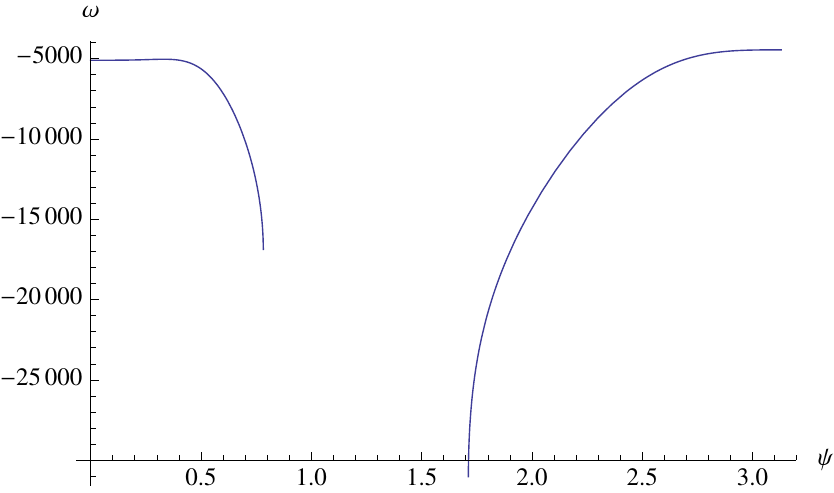} \hskip.3in
\includegraphics[width=0.4\textwidth]
{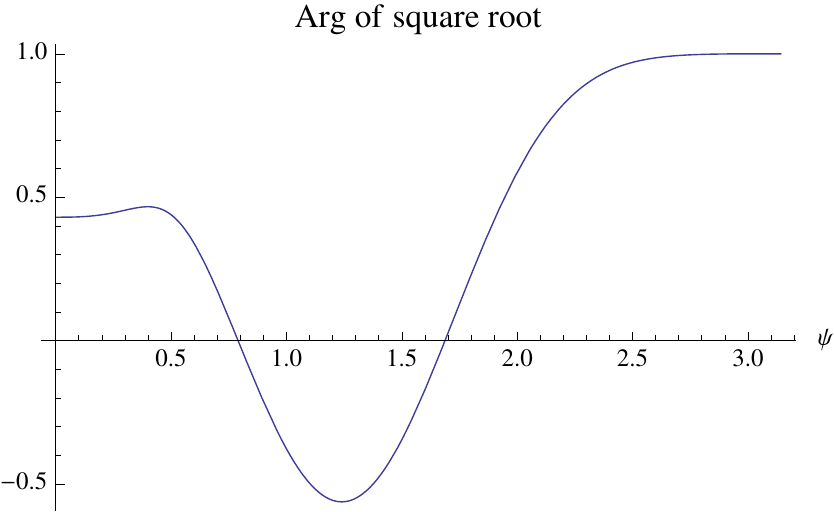}
\caption{The function $\omega$ and the argument of the square root in its formula (\ref{omegasoln}) for $|W_0|=0.02$; note the gap where $\omega$ has no real solution, leading to spontaneous decompactification.
\label{Figomega}}
\end{center}
\end{figure}

\begin{figure}
\begin{center}
\includegraphics[width=0.6\textwidth]
{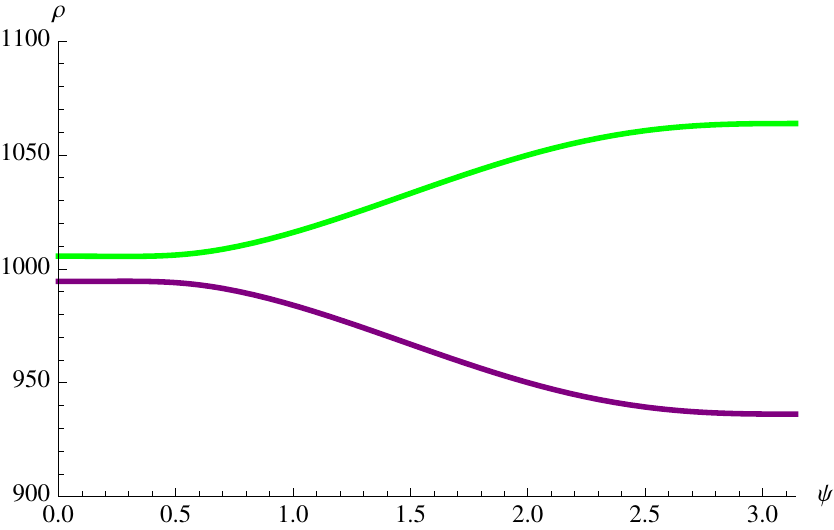} 
\caption{The variation in the K\"ahler modulus $\rho$ for $\mu=2$  (upper) and $\mu=1/2$ (lower).  Values further from unity lead to greater variation as $\log |\mu|$.
\label{Figrho}}
\end{center}
\end{figure}

One can also see that for $|W_0|< 0.025$, the argument of the square root in the solution for $\omega$ goes negative and the decompactification constraint is not satisfied (figure~\ref{Figomega}). Thus we see clearly: very small $|W_0|$ leads to decompactification, larger to perturbative-dominance with allowed decay, larger still to a mixture with vacuum energy near zero and largest of all to nonperturbative dominance with a negative vacuum energy and decay shut down.

Finally, we note that since $\rho$ is of order $10^4$ and $\Delta \rho = M \log \mu$ of order $10^2$, the volume changes on the $10^{-2}$ level during the decay, going from $\rho_i = 1005$ to $\rho_f = 1064$.  We plot the change for $\mu = 2$ as well as $\mu = 1/2$ in figure~\ref{Figrho}, showing both a growth and shrinkage of the compact volume; values of $\mu$ farther away from unity will increase the effect.

\section{Conclusions}

We have demonstrated that the interplay of nonperturbative moduli stabilization and brane/flux annihilation can have a rich variety of effects.
Our results have a number of consequences for the stability and decay of flux compactifications of string theory.  

Stabilization of stacks of antibranes at positive or very near zero cosmological constant, which without nonperturbative physics may decay classically, has obvious phenomenological utility for construction of realistic string vacua.  
The existence of spontaneous decompactification, on the other hand, implies that any vacuum transitions in the early universe mediated by a brane/flux annihilation process are less stable than would otherwise be expected; this places more stringent bounds on the path such a transitioning universe could have taken to avoid ending up in ten dimensions at some point in its history. Moreover, a background that makes many such transitions will have changed its volume with each decay. If the nonperturbative physics is hosted not far down the throat in each case, this can accumulate and cause the volume to grow to a size substantially above the string scale with less than the usual fine tuning required, although if this contribution dominates, another source of positive vacuum energy must be added.

These calculations could be expanded upon in a number of ways.  We have performed the entire analysis assuming the existence of only one K\"ahler modulus; a generic compactification will have many, and it is possible that the interaction between them in addition to their effects on the brane decay could lead to novel effects.  We have not studied the dynamical system including kinetic terms for the K\"ahler modulus as well as the brane coordinate; such a calculation could shed light, for example, on the process of spontaneous decompactification and whether it might be avoided in certain circumstances, for example if the decay occurs sufficiently quickly.  We have not included the perturbative backreaction of the rest of the geometry outside the throat; this is one more possible source of forces on the brane system, as well.   We anticipate, however, that the broad classes of phenomena we have outlined will persist in these more general cases.

\section*{Acknowledgments}

We are grateful for discussions with Shanta de Alwis,  Shamit Kachru, Benjamin Shlaer and Bret Underwood. This work was supported by the DOE under grant DE-FG02-91-ER-40672.

\bibliographystyle{unsrt}

\begin{thebibliography}{99}

\bibitem{KPV}
  S.~Kachru, J.~Pearson and H.~Verlinde,
  ``Brane/flux annihilation and the string dual of a non-supersymmetric field theory,'' 
  JHEP {\bf 0206}, 021 (2002) 
  [arXiv:hep-th/0112197].

\bibitem{Myers}
  R.~C.~Myers,
  ``Dielectric branes,''
  JHEP {\bf 9912}, 022 (1999)
  [arXiv:hep-th/9910053].

  \bibitem{KS}
I.~R.~Klebanov, M.~J.~Strassler,
  ``Supergravity and a confining gauge theory: Duality cascades and
  chiSB-resolution of naked singularities,''
JHEP {\bf 0008}, 052 (2000)
  [arXiv:hep-th/0007191].
  
   \bibitem{Verlinde}
  H.~L.~Verlinde,
  ``Holography and compactification,''
  Nucl.\ Phys.\  B {\bf 580}, 264 (2000)
  [arXiv:hep-th/9906182].

 
  \bibitem{GKP}
  S.~B.~Giddings, S.~Kachru and J.~Polchinski,
  ``Hierarchies from fluxes in string compactifications,''
  Phys.\ Rev.\  D {\bf 66}, 106006 (2002)
  [arXiv:hep-th/0105097].

\bibitem{KKLT}
    S. Kachru, R. Kallosh, A. Linde, S.~P. Trivedi, ``de Sitter vacua in String Theory,''
    Phys. Rev. D, {\bf 68}, 046005 (2003) [arXiv:hep-th/0301240].




\bibitem{Giddings:2003zw}
  S.~B.~Giddings,
  ``The fate of four dimensions,''
  Phys.\ Rev.\  D {\bf 68}, 026006 (2003)
  [arXiv:hep-th/0303031].

\bibitem{Frey:2003dm}
  A.~R.~Frey, M.~Lippert and B.~Williams,
  ``The fall of stringy de Sitter,''
  Phys.\ Rev.\  D {\bf 68}, 046008 (2003)
  [arXiv:hep-th/0305018].


\bibitem{PRZ}
	L.~Pilo, A.~Riotto and A.~Zaffaroni,
	``Old inflation in string theory,''
	JHEP {\bf 0407}, 052 (2004)
	[arXiv:hep-th/0401004v2].

 \bibitem{DKV}
  O.~DeWolfe, S.~Kachru and H.~Verlinde,
  ``The giant inflaton,''
  JHEP {\bf 0405}, 017 (2004) 
  [arXive:hep-th/0403123]


\bibitem{de Alwis:2006cb}
  S.~P.~de Alwis,
  ``Transitions between flux vacua,''
  Phys.\ Rev.\  D {\bf 74}, 126010 (2006)
  [arXiv:hep-th/0605184].

\bibitem{Heckman:2007ub}
  J.~J.~Heckman and C.~Vafa,
  ``Geometrically Induced Phase Transitions at Large N,''
  JHEP {\bf 0804}, 052 (2008)
  [arXiv:0707.4011 [hep-th]].

\bibitem{Sarangi:2007jb}
  S.~Sarangi, G.~Shiu and B.~Shlaer,
  ``Rapid Tunneling and Percolation in the Landscape,''
  arXiv:0708.4375 [hep-th].


\bibitem{Freivogel:2008wm}
  B.~Freivogel and M.~Lippert,
  ``Evidence for a bound on the lifetime of de Sitter space,''
  JHEP {\bf 0812}, 096 (2008)
  [arXiv:0807.1104 [hep-th]].


\bibitem{Ganor}
  O.~J.~Ganor,
  ``A note on zeroes of superpotentials in F-theory,''
  Nucl.\ Phys.\ B {\bf 499}, 55 (1997) [arXiv:hep-th/9612077].



\bibitem{BHK}
  M.~Berg, M.~Haack and B.~Kors,
  ``Loop corrections to volume moduli and inflation in string theory,''
  Phys.\ Rev.\ D {\bf 71}, 026005 (2005)
  [arXiv:hep-th/0404087].
  \\
  M.~Berg, M.~Haack and B.~Kors,
  ``String loop corrections to kahler potentials in orientifolds,''
  JHEP {\bf 0511}, 030 (2005)
  [arXiv:hep-th/0508043].
  

 \bibitem{GM}
  S.~B.~Giddings and A.~Maharana,
  ``Dynamics of warped compactifications and the shape of the warped landscape,''
  Phys.\ Rev.\ D {\bf 73}, 126003 (2006)
  [arXiv:hep-th/0507158]


  \bibitem{Baumann}
  D.~Baumann, A.~Dymarsky, I.~R.~Klebanov, J.~Maldacena, L.~McAllister and A.~Murugan,
  ``On D3-brane potentials in compactifications with fluxes and wrapped
  D-branes,''
  JHEP {\bf 0611}, 031 (2006)
  [arXiv:hep-th/0607050].

\bibitem{DMSU}
  O.~DeWolfe, L.~McAllister, G.~Shiu and B.~Underwood,
  ``D3-brane Vacua in Stabilized Compactifications,''
  JHEP {\bf 0709}, 121 (2007)
  [arXiv:hep-th/0703088].

  
  


\bibitem{BD}
  C.~M.~Brown and O.~DeWolfe,
  ``Nonsupersymmetric brane vacua in stabilized compactifications,''
  JHEP {\bf 0901}, 039 (2009)
  [arXiv:0806.4399 [hep-th]].

\bibitem{I1}
 C.~P.~Burgess, J.~M.~Cline, K.~Dasgupta and H.~Firouzjahi,
  ``Uplifting and inflation with D3 branes,''
  JHEP {\bf 0703}, 027 (2007)
  [arXiv:hep-th/0610320].
  
  
  \bibitem{I3}
  D.~Baumann, A.~Dymarsky, I.~R.~Klebanov, L.~McAllister and P.~J.~Steinhardt,
  ``A Delicate Universe,''
  Phys.\ Rev.\ Lett.\  {\bf 99}, 141601 (2007)
  [arXiv:0705.3837 [hep-th]],\\
    D.~Baumann, A.~Dymarsky, I.~R.~Klebanov and L.~McAllister,
  ``Towards an Explicit Model of D-brane Inflation,''
  JCAP {\bf 0801}, 024 (2008)
  [arXiv:0706.0360 [hep-th]].

  
  \bibitem{KP}
  A.~Krause and E.~Pajer,
   ``Chasing Brane Inflation in String-Theory,''
  JCAP {\bf 0807}, 023 (2008)
  [arXiv:0705.4682 [hep-th]]. 

\bibitem{Panda:2007ie}
  S.~Panda, M.~Sami and S.~Tsujikawa,
  ``Prospects of inflation in delicate D-brane cosmology,''
  Phys.\ Rev.\  D {\bf 76}, 103512 (2007)
  [arXiv:0707.2848 [hep-th]].

    \bibitem{U}
    B.~Underwood,
  ``Brane Inflation is Attractive,''
  arXiv:0802.2117 [hep-th].


  \bibitem{Pajer}
E.~Pajer,
  ``Inflation at the Tip,''
  JCAP {\bf 0804}, 031 (2008)
  [arXiv:0802.2916 [hep-th]].

\bibitem{Chen:2008au}
  F.~Chen and H.~Firouzjahi,
  ``Dynamics of D3-D7 Brane Inflation in Throats,''
  JHEP {\bf 0811}, 017 (2008)
  [arXiv:0807.2817 [hep-th]].

\bibitem{Marchesano:2008rg}
  F.~Marchesano, P.~McGuirk and G.~Shiu,
  ``Open String Wavefunctions in Warped Compactifications,''
  arXiv:0812.2247 [hep-th].

\bibitem{BDKKM}
  D.~Baumann, A.~Dymarsky, S.~Kachru, I.~R.~Klebanov and L.~McAllister,
  ``Holographic systematics of D-brane inflation,'' 
  [arXiv:0808.2811v2 [hep-th]].

\bibitem{CHS}
  H.~Y.~Chen, L.~Y.~Hung and G.~Shiu,
  ``Inflation on an Open Racetrack,''
  arXiv:0901.0267 [hep-th].


  \bibitem{Kuperstein}
  S.~Kuperstein,
  ``Meson spectroscopy from holomorphic probes on the warped deformed
  conifold,''
  JHEP {\bf 0503}, 014 (2005)
  [arXiv:hep-th/0411097].


\bibitem{ABFK1}
   R.~Argurio, M.~Bertolini, S.~Franco and S.~Kachru,
  ``Gauge/gravity duality and meta-stable dynamical supersymmetry breaking,''
  JHEP {\bf 0701}, 083 (2007)
  [arXiv:hep-th/0610212].

\bibitem{ABFK2}
  R.~Argurio, M.~Bertolini, S.~Franco and S.~Kachru,
  ``Metastable vacua and D-branes at the conifold,''
  JHEP {\bf 0706}, 017 (2007)
  [arXiv:hep-th/0703236].

 \bibitem{EKK}
  J.~Evslin, C.~Krishnan and S.~Kuperstein,
  ``Cascading quivers from decaying D-branes,''
  JHEP {\bf 0708}, 020 (2007)
  [arXiv:0704.3484v2 [hep-th]].




\bibitem{Shiu:2008ry}
  G.~Shiu, G.~Torroba, B.~Underwood and M.~R.~Douglas,
  ``Dynamics of Warped Flux Compactifications,''
  JHEP {\bf 0806}, 024 (2008)
  [arXiv:0803.3068 [hep-th]].

\bibitem{Douglas:2008jx}
  M.~R.~Douglas and G.~Torroba,
  ``Kinetic terms in warped compactifications,''
  arXiv:0805.3700 [hep-th].

\bibitem{Frey:2008xw}
  A.~R.~Frey, G.~Torroba, B.~Underwood and M.~R.~Douglas,
  ``The Universal Kaehler Modulus in Warped Compactifications,''
  JHEP {\bf 0901}, 036 (2009)
  [arXiv:0810.5768 [hep-th]].


  
  \bibitem{PS}
  J.~Polchinski and M.~J.~Strassler,
  ``The string dual of a confining four-dimensional gauge theory,''
  [arXiv:hep-th/0003136].

 
 
  \bibitem{Comments}
P. Candelas and X. C. de la Ossa,
   ``Comments On Conifolds,''
    Nucl. Phys. B {\bf 342}, 246 (1990).



\bibitem{Herzog}
  C.~P.~Herzog, I.~R.~Klebanov and P.~Ouyang,
  ``D-branes on the conifold and ${\cal N}=1$,''
  [arXiv:hep-th/0205100].
  \\
  C.~P.~Herzog, I.~R.~Klebanov and P.~Ouyang,
  ``Remarks on the warped deformed conifold,''
  [arXiv:hep-th/0108101].
 
  
 \bibitem{KKLMMT}
  S.~Kachru, R.~Kallosh, A.~Linde, J.~Maldacena, L.~McAllister and S.~P.~Trivedi,
  ``Towards inflation in string theory,''
  JCAP {\bf 0310}, 013 (2003) 
  [arXiv:hep-th/0308055].

\bibitem{OtherFlux}
  S.~Gukov, C.~Vafa and E.~Witten,
  ``CFT's from Calabi-Yau four-folds,''
  Nucl.\ Phys.\ B {\bf 584}, 69 (2000)
  [Erratum-ibid.\ B {\bf 608}, 477 (2001)]
  [arXiv:hep-th/9906070].
\\
  K.~Dasgupta, G.~Rajesh and S.~Sethi,
 ``M theory, orientifolds and G-flux,''
  JHEP {\bf 9908}, 023 (1999)
  [arXiv:hep-th/9908088].
\\
  T.~R.~Taylor and C.~Vafa,
  ``RR flux on Calabi-Yau and partial supersymmetry breaking,''
  Phys.\ Lett.\ B {\bf 474}, 130 (2000)
  [arXiv:hep-th/9912152].

  
\bibitem{AAB}
  O.~Aharony, Y.~E.~Antebi and M.~Berkooz,
  ``Open string moduli in KKLT compactifications,''
  Phys.\ Rev.\  D {\bf 72}, 106009 (2005)
  [arXiv:hep-th/0508080].
    
\bibitem{KMS}
  S.~Kachru, L.~McAllister and R.~Sundrum,
  ``Sequestering in string theory,''
  JHEP {\bf 0710}, 013 (2007)
  [arXiv:hep-th/0703105].

    
\bibitem{RS}
  L.~Randall and R.~Sundrum,
  ``A large mass hierarchy from a small extra dimension,''
  [arXiv:hep-pt/9905221v1]

  
 
\end{thebibliography}

\end{document}